\documentclass[12pt]{article}
\makeatletter
\@addtoreset{equation}{section}
\makeatother

\usepackage[pdftex]{graphicx}
\usepackage{subfigure}
\usepackage{array}
\usepackage{amsmath}

\usepackage{rotating}
\usepackage{longtable}
\usepackage[T1]{fontenc}
\usepackage{mathtools}
\usepackage{natbib}
\usepackage{floatrow}
\usepackage{subfig}

\usepackage{float}
\usepackage{array}
\usepackage{geometry}
\usepackage{color}
\usepackage{lipsum}

\usepackage{diagbox} 
\usepackage{pdfpages}
\usepackage{multicol}
\usepackage[utf8]{inputenc}
\usepackage[english]{babel}
\usepackage{mathtools}
\usepackage{bbm}
\usepackage{amssymb}
\usepackage{amsthm}
\usepackage{hyperref}
\usepackage{bm}
\usepackage{mathtools}
\usepackage{fancybox}
\usepackage{multirow}
\usepackage{authblk}
\usepackage{algorithm}
\usepackage{algpseudocode}
\usepackage{setspace}
\usepackage{caption}
\usepackage{fancybox}

\usepackage{adjustbox}
\usepackage[utf8]{inputenc}
\usepackage{tikz}

\allowdisplaybreaks

\usetikzlibrary{shapes.geometric,arrows}
\tikzstyle{startstop} = [rectangle, rounded corners, minimum width=3cm, minimum height=1cm,text centered, draw=black]
\tikzstyle{io} = [trapezium, trapezium left angle=70, trapezium right angle=110, minimum width=3cm, minimum height=1cm, text centered, draw=black]
\tikzstyle{process} = [rectangle, minimum width=3cm, minimum height=1cm, text centered, draw=black]
\tikzstyle{decision} = [diamond, minimum width=3cm, minimum height=1cm, text centered, draw=black]
\tikzstyle{arrow} = [thin,->,>=stealth]

\newlength\szg
\newcommand\quan[1]{%
\settoheight\szg{#1}%
\tikz[baseline]{\pgfmathparse{
    ifthenelse(#1 < 10, 1, ifthenelse(#1 < 100, 0.75, 0.5))
}
\let\hfs\pgfmathresult
\node at (0,\szg/2) {\makebox[0em][c]{\scalebox{\hfs}[1]{#1}}};
\draw (0,\szg/2) circle (\szg/2+0.35ex);
}}

\makeatletter
\newcommand*{\rom}[1]{\expandafter\@slowromancap\romannumeral #1@}
\makeatletter
\def\BState{\State\hskip-\ALG@thistlm}
\makeatother
\usepackage{xcolor}
\renewcommand{\raggedright}{\leftskip=0pt \rightskip=0pt plus 0cm}
\title{Filtrated Common Functional Principal Components Analysis of Multi-group Functional data}
\author[1]{Shuhao Jiao\thanks{shjiaoqd@gmail.com}}
\author[2]{Ron D.\ Frostig\thanks{rfrostig@uci.edu}}
\author[1]{Hernando Ombao\thanks{hernando.ombao@kaust.edu.sa}}

\affil[1]{Statistics Program,  KAUST, Saudi Arabia}
\affil[2]{Department of Neurobiology and Behavior, UC Irvine, USA}
\date{}
\allowdisplaybreaks
\begin{document}
	\maketitle			
	\setlength\parindent{0pt}
	\setlength{\parskip}{1em}
	\theoremstyle{definition}
	\newtheorem{theorem}{Theorem}
	\newtheorem{lemma}{Lemma}
	\newtheorem{assumption}{Assumption}
	\newtheorem{prop}{Proposition}
	\newtheorem{definition}{Definition}
	\newtheorem{remark}{Remark}

\begin{abstract}
Local field potentials (LFPs) are signals that measure electrical activity in localized cortical regions from implanted tetrodes in the human or animal brain. The LFP signals are curves observed at multiple tetrodes which are implanted across a patch on the surface of the cortex. Hence, they can be treated as multi-group functional data, where the trajectories collected across temporal epochs from one tetrode are viewed as a group of functions. In many cases, multi-tetrode LFP trajectories contain both global {variation patterns} (which are shared in common to all groups, due to signal synchrony) and isolated {variation patterns}  (common only to a small subset of groups), and such structure is very informative to the analysis of such data.  Therefore, one goal in this paper is to develop an efficient procedure that is able to capture and quantify both global and isolated features. We propose a novel tree-structured functional principal components (filt-fPC) analysis through finite-dimensional functional representation -- specifically via filtration. A major advantage of the proposed filt-fPC method is the ability to extract the components that are common to multiple groups (or tetrodes) in a flexible "multi-resolution" manner and simultaneously preserve the idiosyncratic individual components of different tetrodes. The proposed filt-fPC approach is highly data-driven and no "ground-truth" model pre-specification is needed, making it a suitable approach for analyzing multi-group functional data that is complex. In addition, the filt-fPC method is able to produce a  parsimonious, interpretable, and efficient low dimensional representation of multi-group functional data with orthonormal basis functions. Here, the proposed filt-fPCA method is employed to study the impact of a shock (induced stroke) on the synchrony structure of the rat brain. The proposed filt-fPCA is a general approach that can be readily applied to analyze other complex multi-group functional data, such as multivariate functional data, spatial-temporal data and longitudinal functional data. 

\noindent{\bf Key words}: Functional principal components, Community detection, Dimension reduction, Multi-group functional data, Network filtration, Supervised learning, Weighted network

\end{abstract}

\section{Introduction}
\subsection{Data description and statistical challenges}
\begin{figure}[ht]
\center
\includegraphics[scale=0.3]{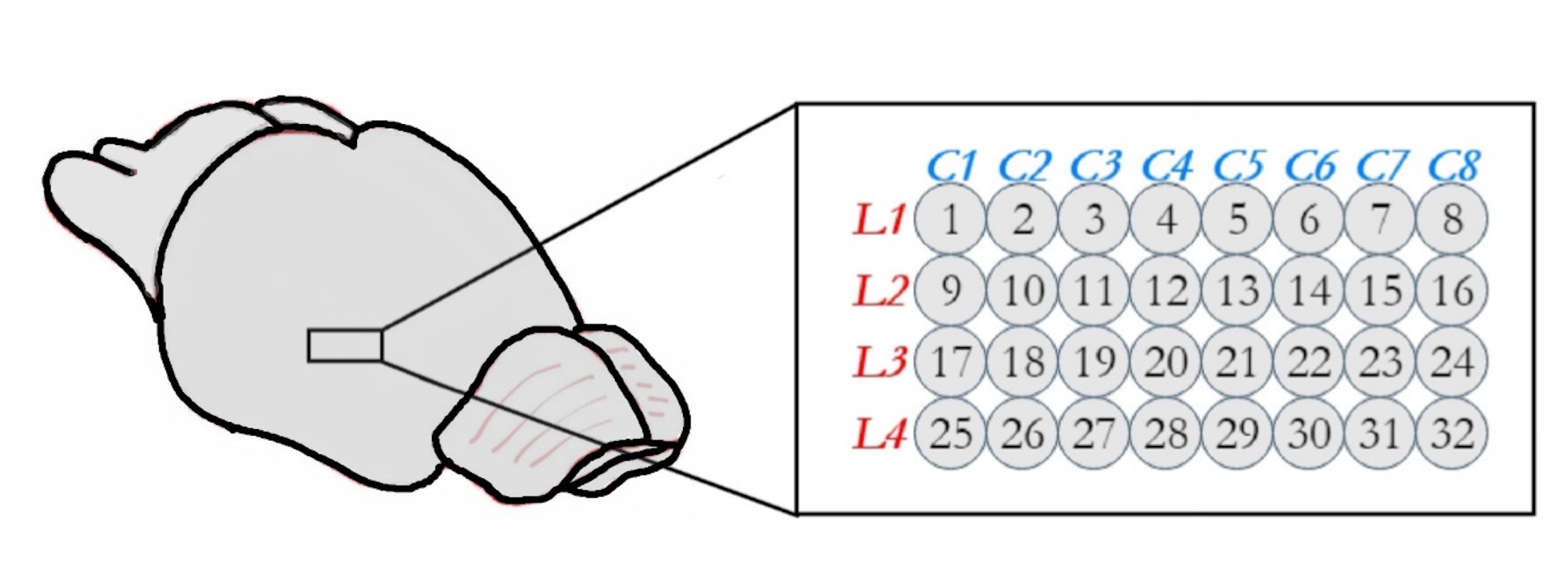}
\caption{Placement of 32 tetrodes}
\label{placement}
\end{figure}

This work is motivated by a neuroscience study conducted by co-author, Ron D.~Frostig, where the goal is to investigate the impact of an extreme shock (such as a stroke) on the functional organization of the rat brain. In the experiment  described in Wann (2017) \cite{wann2017large},  ischemic stroke was simulated by clamping the medial cerebral artery of a rat. Brain activity was continuously monitored over several hours (both pre-occlusion and post-occlusion/clamping) through the local field potential (LFP) recordings from 32 implanted micro-tetrodes (see Figure \ref{placement}). In this set-up, two temporal phases of the LFP recordings were considered: pre-occlusion and post-occlusion. 

Since the mean curves are always close to zero for brain signals, one goal in dimension reduction of such functional data is to identify the commonality of covariance structures across different tetrodes, which is highly associated with synchrony in this neuroscience project. Here, we say "covariance structure" instead of "covariance operator" because the major interest is the variation pattern rather than the variation level. As the covariance structure of different tetrodes can be quite similar due to the synchrony phenomenon, it is possible to obtain a more parsimonious representation -- via the filt-fPC -- by employing common components across different tetrodes. Such common components are informative to the synchrony structure of multi-tetrode LFPs.

Note that signal synchrony is just a special reason that leads to the feasibility of employing common principal components. Two groups of functions can share common principal components as long as they share similar covariance structure. In this paper, the developed new method has the ability to extract the common components for general multi-group functional data.
 
\begin{figure}[ht]
\center
\includegraphics[scale=0.7]{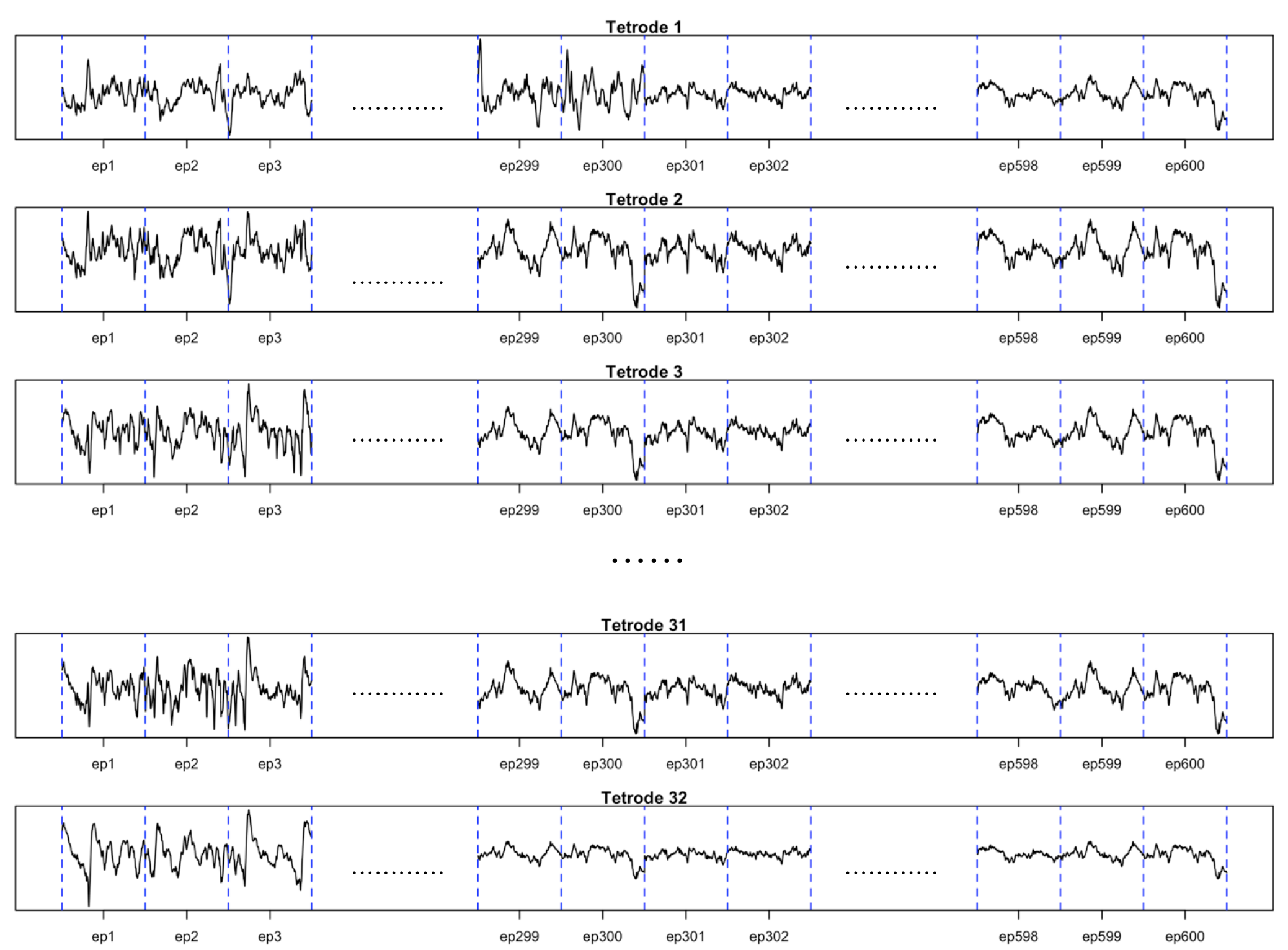}
\caption{{\color{black}Diagram of LFPs. The stroke was induced at the end of the 300-th epoch (ep300), each epoch has 1000 recordings observed over 1 second.}}
\label{diagram}
\end{figure}

In this paper, the focus will be on the 10-minute period around the time of occlusion: \ 5 minutes immediately prior to occlusion of the medial cerebral artery (pre-occlusion)  and 5 minutes during the post-occlusion phase.  The goal is \underline{not} to classify the signals into pre-occlusion vs post-occlusion phases. Rather, the aims are (i.) to extract the {\it intrinsic} functional structure that is present {during each phase respectively}; and (ii.) to identify and quantify the differences in the functional structure between the two phases. To conduct our analysis, the LFP data was segmented into separate 1-second epochs and thus, each of the pre-occlusion and post-occlusion phases consists of 300 epochs for each tetrode. See Figure \ref{diagram} for the diagram of the LFP data. 
It is noted that some tetrodes produce trajectories displaying similar variation patterns due to signal synchrony, and the level of synchrony is not the same across tetrodes. It is potentially critical to check how synchrony is affected by ischemic stroke for understanding how the temporal coordination in the function of large-scale brain networks are associated with the functional impairments caused by ischemia. 


\subsection{Existing fPCA methods for multi-group functional data}
Functional data analysis is an active area  primarily driven by its wide range of potential applications. Due to the infinite dimensionality of functional data, one of the fundamental techniques employed in the analysis is dimension reduction. Functional principal component analysis (fPCA), as described in Ramsay and Silverman (2004) \cite{ramsay2004functional}, is a widely-applied dimension reduction technique for univariate functional data analysis because it gives the optimal approximation of functions (with respect to the integrated squared error of reconstruction) and at the same time yields results that are interpretable in the sense that it captures the principal directions of variation. {More work on fPCA include, but are not limited to, Hall and Hosseini-Nasab (2006) \cite{hall2006properties1}, Hall, M\"uller and Wang (2006) \cite{hall2006properties2}, Yao and Lee (2006) \cite{yao2006penalized}, Yao (2007) \cite{yao2007functional}, Jiang and Wang (2010) \cite{jiang2010covariate}, Bali et al.~(2011) \cite{bali2011robust}}.

In many experiments, multiple groups of functional trajectories are collected for a sample of experimental units. There is a need for statistical methods to account for the common variation patterns of multiple groups of functions, which, in our application, {are present} across tetrodes in both the pre-occlusion and post-occlusion phases. However, there are only a few methods that are appropriate for this type of functional data. We now describe these methods and discuss the advantages of filt-fPCA.

The naive approach of applying ordinary univariate fPCA by combining all groups into one big group may not be suitable for multi-group functional data, especially when there is substantial variation in the covariance functions of different groups.  Consequently, the fPCs obtained are not guaranteed to be capable of explaining the major variation in {\it all} groups.  At the other end of this spectrum, employing group-wise fPCs to each group {separately} can be problematic as well, because this leads to a large number of distinct fPCs when the number of groups is large and cannot reveal the connection between different groups. Several methods of functional principal component analysis for multivariate functional data have been recently developed. A multilevel functional principal components method is proposed in Di et al.~(2009) \cite{di2009multilevel} which explains the variation both within and between different groups of functions. This was extended to sparse sampled multilevel functional data in Di, Crainiceanu and Jank (2014) \cite{di2014multilevel}.  In  Kayano and Konishi (2009) \cite{kayano2009functional},  functional principal component analysis has been developed for multivariate functions with Gaussian-shape basis. Berrendero, Justel and Svarc (2011) \cite{berrendero2011principal} proposed the multivariate principal component with functional scores. A framework for longitudinal functional principal components in Greven et al.~(2011) \cite{greven2011longitudinal}  combines the covariance of within-subject and between-subject components. A two-step fPCA method for longitudinal functional data has been developed in Chen and M\"uller (2012) \cite{chen2012modeling}  where the fPCA is implemented according to different longitudinal indexes, and the resulting principal components vary as the longitudinal index changes. An extension is proposed in Chen, Delicado Useros and M\"uller (2017) \cite{chen2017modelling} which gives a framework with a more parsimonious fPC representation -- marginal fPC representation. Chiou, Chen and Yang (2014) \cite{chiou2014multivariate} and Jacques and Preda (2014) \cite{jacques2014model} proposed the multivariate functional principal component analysis (MfPCA) {which describe the variation pattern of multivariate functional data, and Happ and Greven (2018) \cite{happ2018multivariate} extended the MfPCA for functions defined over different domains. Another related track of research is common principal component analysis (CPCA, see e.g., Flury (1984) \cite{flury1984common}, Benko, H\"ardle and Kneip (2009) \cite{benko2009common} and partial common principal component analysis (PCPCA, see e.g., Flury (1987) \cite{flury1987two}, Schott (1999) \cite{schott1999partial}, Wang et al.~(2019) \cite{wang2019semiparametric}). In  Crainiceanu et al.~(2011) \cite{crainiceanu2011population}, a population value decomposition (PVD) procedure is proposed. Moreover, Lock et al.~(2013) \cite{lock2013joint} and Feng et al.~(2018) \cite{feng2018angle} proposed JIVE and AJIVE procedures to extract common and individual components for multi-block data. While some of the existing methods (i.e., CPCA and PCPCA) and the proposed filt-fPCA share the same goal of finding a common "representation" of multi-group functional data, the principles and algorithms are different. In CPCA and PCPCA, all groups are assumed to share the same set of common principal components. Such assumption can be overly restrictive practically. {Comparatively, the filt-fPCA aims to find a reasonable common fPC system, and has the ability of extracting "multi-resolution" common features of different groups, {\color{black}and thus the number of (filtrated) common functional principal components is allowed to vary across groups. In addition, compared to the existing methods, filt-fPCA has the ability to produce flexible and efficient (in functional reconstruction) fPC commonality structures, which identifies the global (common across groups) and local (group-specific) structure, and works for both balanced and unbalanced designs. }
\subsection{Filtrated functional principal component analysis}
\label{s1.3}
In this article, we propose a new procedure to extract common fPCs (filt-fPCs) that produces {a low-dimensional representation for multi-group functional data}. The filt-fPCA method builds on the idea of filtration in network analysis and extracts the common principal components via {multi-layer} filtrations, {\color{black}and we will employ the filt-fPCA method to extract the commonality structure of multi-tetrode epochs for both pre-occlusion and post-occlusion phases, and compare the structure between the two phases.} 

The {\it fundamental philosophy} of filt-fPCA is that, as the common components are extracted from different groups of functions ({\color{black}in our application, the functions are the epoch trajectories, and the epochs collected from the same tetrode are viewed as a group}) with an increasing number of layers, the variation patterns of residuals across different groups tend to be more idiosyncratic. {\color{black}Define a {\it community} as a collection of multiple groups of functions. In filt-fPCA, the most common components are first extracted, and in this step, all groups of functions are clustered into some (comparatively) large communities. Then we split the large communities into smaller ones, and the second common components are extracted for each of the communities. This splitting \& extraction procedure is repeated till some ending conditions are satisfied.}  
Accordingly, we propose to employ a tree-structured common filt-fPC system as displayed in Figure \ref{tree}. As the filtration goes, all groups are layer-wisely split into smaller communities, where "higher-resolution" (more group-specific) common components are to be extrated. The communities in a layer are always nested in, or identical to, the communities of the previous layer. The filt-fPCs in the first layer of the tree structure capture the most common variation pattern, and the filt-fPCs in the layers formed later in the tree {pertain to} more idiosyncratic variation patterns which are shared by a fewer number of groups. For each community, one common filt-fPC is extracted for all groups of functions belonging to that community. Clearly, a primary step before obtaining filt-fPCs is to find a decent {\it community structure} which includes the communities in all incorporated layers. {\color{black}Note that the commonality structure proposed in PCPCA is just a special case of our proposed tree structure. Specifically, in PCPCA, in the layers pertaining to commonality, all groups are assigned to the same community, and in the layers pertaining to idiosyncrasy, all groups are separated from each other. In addition, group-wise ordinary fPCs can also be viewed as a special case of filt-fPCs, and in this special case, all groups are separated in all layers.}

\begin{figure}[ht]
\center
\includegraphics[scale=0.4]{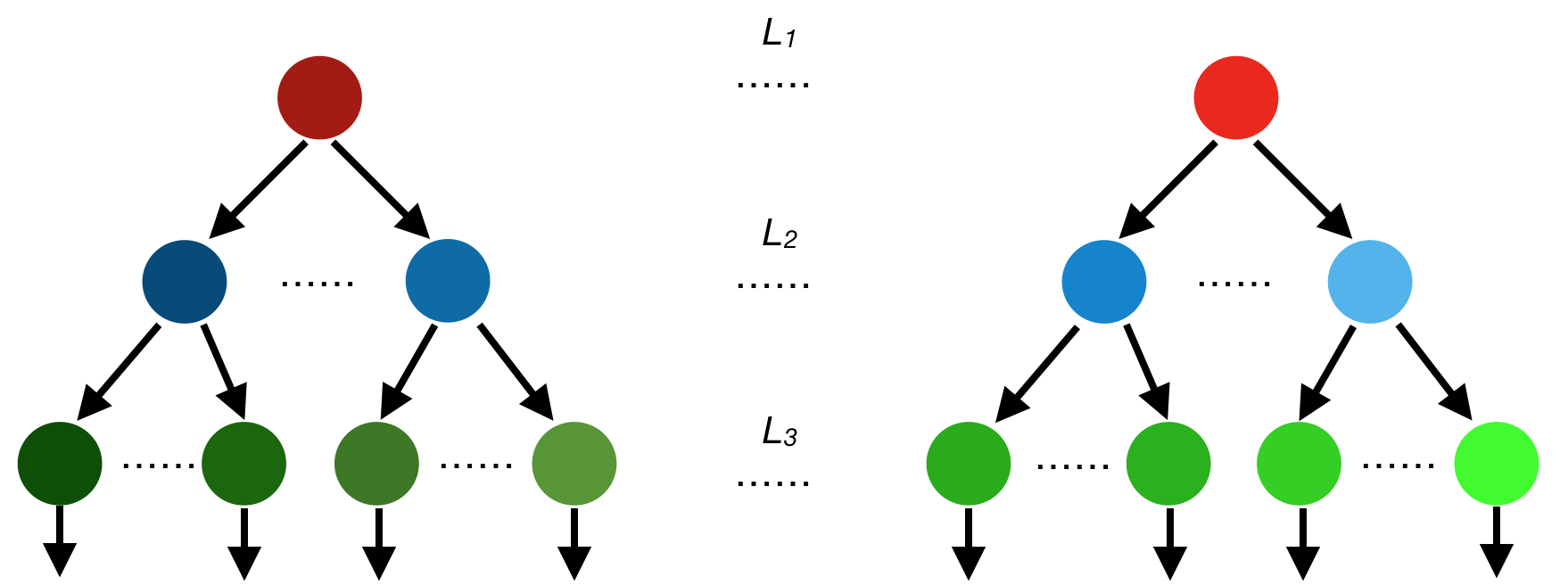}
\caption{Hierarchical tree structure of filt-fPCs. Here, $L_j$ signifies the $j$-th layer of filtration. Different communities are represented by different colored solid circles. Note that there can be multiple trees since the groups may be partitioned into multiple communities in the first layer (one tree per first-layer community).}
\label{tree}
\end{figure}

In summary, the proposed filt-fPCA has the following advantages: 
1.) The {\color{black}searching of the tree-structure} is data-driven without constraints of model specification, and no prior knowledge is needed to implement the method, making it suitable for complex multi-group functional data.
2.) The method produces functional principal components that simultaneously reveal both the common and individual features of various groups of functions in a "multi-resolution" manner, which enables sophisticated analysis and provides more comprehensive explanation of multi-group functional data. 
3.) The method produces basis functions that are orthonormal for each group and thus greatly reduces computational burden in the sequential analysis. 
4.) The proposed method is applicable to both balanced and unbalanced design of multi-group functional data.

The rest of the paper is organized as follows. In Section~\ref{s2}, some preliminaries of functional data are introduced. In Section~\ref{s3}, we develop the filt-fPCA procedure, including network filtration, community detection and selection, and computing of filtrated functional principal component. Section~\ref{s4} presents some simulation results. Section~\ref{s5} presents the real data analysis on the local field potentials of rat brain activity. Conclusions and summaries are made in Section~\ref{s6}. Technical proofs, pseudocode, and additional figures can be found in the supplementary material.

\section{Preliminaries}
\label{s2}
Denote $X(t)\in L^p_H=L_H^p(\Omega,\mathcal{A},\mathbb{P})$ to be such that, for some $p>0$, a $H$-valued function $X(t)$ satisfies $E\{\|X(t)\|^p\}<\infty$. Here $\|\cdot\|$ is the $\ell^2$-norm defined for elements in $H$. In what follows, all trajectories $\{X_n(t)\colon n\in\mathbb{N}\}$ are assumed to be functions defined in the Hilbert space $L^2[0,1]$, where the inner product is defined as $\langle x,y \rangle=\int_0^1x(t)y(t)dt$, and the norm is defined as $\|x\|^2=\int_0^1x(t)^2dt<\infty$. Suppose that $X(t)\in L_H^1$, then the mean function is defined to be $\mu(t)=E\{X(t)\}$.  Moreover, if $X(t)\in L_H^2$, then the covariance operator is defined to be $\Gamma\colon L^2[0,1]\to L^2[0,1]$ by $\Gamma(\cdot)=E\{\langle X-\mu,\cdot\rangle (X-\mu)(t)\}$. 

By the Mercer's theorem, the following expression holds for the covariance operator $\Gamma(\cdot)$, $\Gamma(\cdot)=\sum_{j=1}^{\infty}\theta_j\langle \nu_j,\cdot\rangle \nu_j(t),$ where $\{\theta_j\colon j\in\mathbb{N}_+\}$ are the positive eigenvalues (in strictly descending order) and $\{\nu_j(t) \colon j \in\mathbb{N_+}\}$ are the corresponding normalized eigenfunctions, so that $\Gamma(\nu_j) = \theta_j\nu_j$ and $\|\nu_j\| = 1$. Here, $\{\nu_j(t) \colon j \in\mathbb{N}_+\}$ forms a sequence of orthonormal bases for $L^2[0, 1]$. Let $\{X_n(t)\colon n\in\mathbb{N}\}$ be a sequence of random functions with mean function $\mu(t)$ and covariance operator $\Gamma(\cdot)$. By the Karhunen-Lo\`eve theorem, under mild conditions, $X_n(t)$ admits the representation $X_n(t) = \mu(t)+\sum_{j=1}^{\infty}\langle X_n-\mu, \nu_j\rangle \nu_j(t)$. Suppose that there are $N$ samples $X_1(t),\ldots,X_N(t)$, then the estimator of $\mu(t)$ is $\hat{\mu}(t)=N^{-1}\sum_{n=1}^N X_n(t),$ and the estimator of the covariance operator is given by $\widehat{\Gamma}(\cdot)=N^{-1}\sum_{n=1}^N\langle X_n - \hat{\mu}, \cdot\rangle(X_n - \hat{\mu})(t).$

Suppose that there are $G$ groups of functions $\{X_{vn}(t)\colon v=1,\ldots,G,\ n\in\mathbb{N}\}$, where $X_{vn}(t)$ is the $n$-th function in group $v$. Without loss of generality, assume the mean function of each group $v$ is zero. Then for a community $\mathcal{K}$ including some of the groups, the {leading} common filtrated functional principal component is defined as $\arg\min_{\|\phi\|=1}\sum_{v\in \mathcal{K}}f_vE\|X_{vn}-\langle X_{vn},\phi\rangle X_{vn}\|^2,$ where $\{f_v\colon v=1,\ldots,G\}$ specifies the weight of different groups.  A large value of $f_v$ leads to higher influence of the corresponding group on the common filt-fPC.
\begin{remark}
$\{f_v\colon v=1,\ldots,G\}$ should be specified according to specific needs. As a special case, if all groups are equally important, one appropriate  way to specify $\{f_v\colon v=1,\ldots,G\}$ is $f_v=1/\sum_{j\ge1}\theta_{vj}$, where $\theta_{vj}$ is the $j$-th eigenvalue of the covariance operator of the $v$-th group.
\end{remark}
The following proposition provides some guidance to find the leading filt-fPC for a given community $\mathcal{K}$.
\begin{prop}
\label{prop1}
Suppose that, for each $v\in\mathcal{K}$, $X_{vn}(t)\in L^2_H$. The minimizer of $\sum_{v\in \mathcal{K}}f_vE\|X_{vn}-\langle X_{vn},\phi\rangle\phi\|^2$ under $\|\phi\|=1$ is the eigenfunction corresponding to the largest eigenvalue of $\sum_{v\in \mathcal{K}}f_v\Gamma_v(\cdot)$, where $\Gamma_v(\cdot)=E\{\langle X_{vn},\cdot\rangle X_{vn}\}$. 
\end{prop}
Proposition \ref{prop1} provides a blueprint for obtaining the filt-fPCs given a community. In practice, we replace $\sum_{v\in \mathcal{K}}f_v\Gamma_v(\cdot)$ with its empirical version $\sum_{v\in \mathcal{K}}f_v\widehat{\Gamma}_v(\cdot)$. 

\section{Filtrated Common Functional Principal Component} 
{\color{black}In this section, we will develop the conception of filt-fPCA and illustrate the implementation details. The section can be segmented into two parts: 1) the definition and estimation of filt-fPCs (Section \ref{s3.0} and \ref{s3.5}) and 2) the detection and selection of the tree-structured communities (Section \ref{s3.2} and \ref{s3.3}).}
\label{s3}
\subsection{The filt-fPC representation}
\label{s3.0}
Recall that the primary aim is to find the principal components that jointly explain the variation of multiple groups in a filtrating manner, i.e., to obtain the common filt-fPCs for each of the tree-structured communities, and a community is defined to be a set of groups where one common filt-fPC is employed for all the groups in that set. The key idea of filt-fPCA is to represent $\{X_{vn}(t)\colon  v=1,\ldots,G,\ n=1,\ldots,N_v\}$ in the following form,
\begin{align}
\label{filtrep}
X_{vn}(t)=\mu_ v(t)+\sum_{d=1}^\infty\langle X_{vn}-\mu_ v,\phi^{(c_{ v,d})}_d\rangle \phi^{(c_{ v,d})}_d(t),
\end{align}
where $c_{ v,d}$ is the community index of group $v$ in the $d$-th layer of filtration, {$\{\phi_d^{(c_{v,d})}(t)\colon d\ge1\}$ are the filt-fPCs of group $v$} (group $v$ and group $v'$ share the same filt-fPC in dimension $d$ if $c_{v,d}=c_{v',d}$) satisfying $\langle\phi_{d}^{(c_{v,d})},\phi_{d'}^{(c_{v,d'})}\rangle=0$ as $d\ne d'$ and $\|\phi_{d}^{(c_{v,d})}\|=1$, and $\mu_{v}(t)$ is the mean function of the $v$-th group, which is zero in our project since LFPs always oscillate around the zero-line. $\{\langle X_{vn}-\mu_ v, \phi^{(c_{ v,d})}_d\rangle\colon d\ge1\}$ are the filt-fPC scores of $X_{vn}(t)$. 

\begin{remark}
Note that, since functional data is infinite-dimensional, a community structure can include up to infinitely many layers, and there exists at least one layer of  structure shared across infinitely many consecutive layers, but in practice only a finite number of layers are incorporated since it is very hard to implement statistical analysis in an infinite-dimensional space. More details will be discussed in Section \ref{s3.3}.
\end{remark}

\subsection{The estimation of filt-fPC}
\label{s3.5}
{\color{black}Given the tree-structured communities, the estimation of filt-fPCs is easy-to-implement. Denote $\mathcal{K}_{d1},\mathcal{K}_{d2},\ldots$ as the communities in the $d$-th layer, and 
$$R^{(d)}_{vn}(t)=X_{vn}-\sum_{j=1}^{d}\langle X_{vn},\hat{\phi}_j^{(c_{v,j})} \rangle\hat{\phi}_j^{(c_{v,j})}$$ as the projection residual in the $d$-th layer, and $\Gamma_{v}^{(d)}(\cdot)$ as the covariance operator of $R^{(d)}_{vn}(t)$. As a special case, $R^{(0)}_{vn}(t)=X_{vn}(t)$. The common filtrated fPC of groups in $\mathcal{K}_{di}$ is defined as the maximizer of $$h(\phi(t))=\left\langle\sum_{v\in\mathcal{K}_{di}}f_v^{(d-1)}\Gamma^{(d-1)}_v(\phi),\phi\right\rangle$$ over all normalized function $\phi(t)$, which measures the weighted sum of variability of groups in $\mathcal{K}_{di}$ explained by $\phi(t)$. 
The maximizer of the above quantity, denoted by $\hat{\psi}_{di}$, is obtained as:
$$\hat{\psi}_{di}=\arg\max\limits_{\|\phi\|=1}\sum_{v\in \mathcal{K}_{di}}\sum_{n=1}^{N_v}\frac{f^{(d-1)}_v}{N_v}\langle R^{(d-1)}_{vn},\phi\rangle^2,\qquad \text{for}\  v\in \mathcal{K}_{di},\ i\ge1,$$
and $\hat{\phi}_d^{(c_{v,d})}=\hat{\psi}_{di}$ if $v\in\mathcal{K}_{di}$. As illustrated in Proposition~\ref{prop1}, the maximizer of the objective function 
$\sum_{v\in \mathcal{K}_{di}}\sum_{n=1}^{N_v}N_v^{-1}f^{(d-1)}_v\langle R^{(d-1)}_{vn},\phi\rangle^2$ is the first eigenfunction of the operator $\sum_{v\in \mathcal{K}_{di}}f^{(d-1)}_v\widehat{\Gamma}^{(d-1)}_v(\cdot),$
where 
$$\widehat{\Gamma}^{(d-1)}_v(\cdot)=N_v^{-1}\sum_{n=1}^{N_v}\{R^{(d-1)}_{vn}\langle R^{(d-1)}_{vn},\cdot\rangle\}.$$ 
Then, given the empirical filt-fPCs $\{\hat{\phi}_d^{(c_{v,d})}\colon d\ge1\}$ and total number of layers $D$, the reconstruction of $X_{vn}(t)$ is
$$X_{vn}(t)\approx\mu_v(t)+\sum_{d=1}^D\langle X_{vn}-\mu_v,\hat{\phi}_d^{(c_{v,d})}\rangle\hat{\phi}_d^{(c_{v,d})}(t).$$
The selection of $D$ and the community structure will be discussed in Section \ref{s3.3}.}
\begin{remark}
If all groups are equally important in all layers, a proper way to specify $f_v^{(d)}$ is $f_v^{(d)}=1/\sum_{j\ge1}\theta^{(d-1)}_{vj}$, where $\theta^{(d-1)}_{vj}$ is the $j$-th eigenvalue of the covariance operator of $R_{vn}^{(d-1)}$. {\color{black}Without the scaling step, the groups with higher variation are more influential on the common filt-fPCs}. 
\end{remark}

\begin{prop}
\label{prop2}
The filtrated principal components $\{\phi_d^{(c_{v,d})}(t)\colon d\ge1\}$ are orthonormal for any $v\ge1$. 
\end{prop}

This proposition is important in the sequential analysis, e.g., in the finite-dimensional projection of functional linear models, orthonormality avoids cross terms and hence leads to a concise finite-dimensional representation.

\subsection{Weighted network, filtration, and community detection}
\label{s3.2}
A preliminary step is to evaluate the similarity of the covariance structures. To motivate our approach,  we first develop the notion of similarity through a weighted network model. A weighted network is a triple $(N,E,\bm{\omega})$, where $N$ is the node set representing groups, $E$ is the edge set, and $\bm{\omega}$ is the set of edge weights. In our application, the nodes represent tetrodes; edges can be viewed as existence of synchrony between different regions of brain, which is complete (all pairs of nodes are connected) at the beginning of filtration since we have no prior knowledge that some pairs of regions are not synchronized at all; weights represent the similarity of covariance structures, and also reveal the level of synchrony. A {\it small} value of edge weight indicates {\it similar} variation pattern of the adjacent nodes (higher level of synchrony of the corresponding two tetrodes). 

The principle is that, if two nodes are connected by an edge, then the functions in these two groups are considered to share some similar variation patterns, and thus {likely} share some common filt-fPCs. The edge weight $\omega$ should be a reasonable measure of similarity of variation pattern. Notationally, denote the weight of edge adjacent to nodes $i$ and $j$ as $\omega_{ij}$, and {we propose to set} $\omega_{ij}=\|\mathcal{C}_i-\mathcal{C}_j\|_\mathcal{S},$ where $\mathcal{C}_i$ is the scaled covariance operator of the $i$-th group, defined as $\mathcal{C}_i=\Gamma_i/\sum_{j\ge1}\theta_{ij}$, $\theta_{ij}$ is the $j$-th eigenvalue of $\Gamma_i$, and $\|\cdot\|_\mathcal{S}$ denotes the Hilbert-Schmidt norm. 

{Network filtration is a multi-thresholding framework for displaying the dynamic pattern of how network features change over different thresholds.} 
Here, we propose to specify a sequence of positive thresholds $\{\tau_d\colon d\ge1\}$ {in non-ascending order ($\tau_1\ge\tau_2\ge\tau_3\ge\cdots$)}. For each $d$, we eliminate the edges of which the weights are greater than the threshold $\tau_d$. Corresponding to the thresholds, a sequence of nested network are obtained after the edge truncation $\{(N,E_d,\bm{\omega})\colon d\ge1\}$, where $E_1\supseteq E_2\supseteq\ldots$. Next apply community detection algorithm to separate the nodes into different disjoint communities for each $(N,E_d,\bm{\omega})$. We illustrate the idea of filtration and community in Figure~\ref{f3}. In the first and second filtration layer, the nodes are separated into one and three communities, and thus one and three distinct filt-fPCs are obtained in the first two layers. The last filtration eliminates all the edges, and the filt-fPCs, starting from the third layer, pertain to the idiosyncratic variation pattern of each individual group. The community detection algorithm applicable here is not unique. We introduce our proposal in the supplementary materials. Since the focus of this paper is not to develop a new community detection algorithm, we do not introduce more details.
\begin{figure}[ht]
\vspace{5mm}
    \includegraphics[scale=0.31]{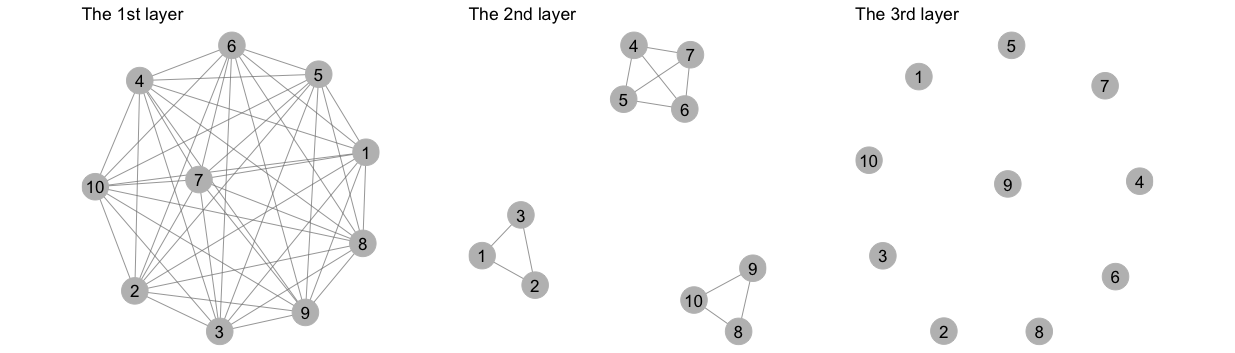}
    \caption{Network filtration and the community structure of the first three layers.}
    \label{f3}
\end{figure}

\subsection{Selection of community structure}
\label{s3.3}
\subsubsection{Generalized information criterion}
Motivated by the generalized information criterion (GIC, see e.g., Nishii (1984) \cite{nishii1984asymptotic} and Zhang, Li and Tsai (2010) \cite{zhang2010regularization}), we propose a penalized criterion, which follows the format: measure of model fit \ $+$ \  tuning parameter $\times$ measure of model complexity. The idea of adding this penalty comes from the fact that, if the groups are split into too many communities, the resulting filt-fPCs will very likely fail to capture the commonality of covariance structures well, although they are efficient in functional reconstruction.

For a given community structure $\bm{C}_\alpha$ ({$\alpha$ is the index of community structure}) and the corresponding empirical filt-fPCs $\{\hat{\phi}_{\alpha,d}^{(c_{v,d})}(t),~d\ge1,~v=1,\ldots,G\}$, we define the GIC value to be
\begin{equation}
\mbox{GIC}(\bm{C}_{\alpha,1:D})=-\sum_{v=1}^GN_v^{-1}f_v\left\{\sum_{n=1}^{N_v}\sum_{d=1}^D\langle X_{vn},\hat{\phi}^{(c_{v,d})}_{\alpha,d}\rangle^2\right\}+\lambda_N(\aleph \bm{C}_{\alpha,1:D}),
\label{GIC}
\end{equation}
where $\aleph$ signifies the cardinality (the total number of communities), and $\bm{C}_{\alpha,1:D}$ represents the first $D$ layers of community structure $\bm{C}_\alpha$. Given $D$, the optimal community structure is defined as the minimizer of $\mbox{GIC}(\bm{C}_{\alpha,1:D})$. 

To illustrate the asymptotic property of the selected community structure, we now introduce the following concept. Let $\{\phi_{vd}\colon d\ge1\}$ be the group-wise ordinary fPCs of the $v$-th group, and the following concept quantifies the difference between the filt-fPCs and group-wise ordinary  fPCs.
\begin{definition}[$\tau$-oracle community structure]
A community structure $\widetilde{\bm{C}}_\tau$ is termed a $\tau$-oracle community structure if
$$\sigma_{\alpha,D}\coloneqq E\left\{\sum\limits_{v=1}^G\sum\limits_{d=1}^D\left(\langle X_{vn},\phi_{vd}\rangle^2-\langle X_{vn},{\phi}^{(c_{v,d})}_{\alpha,d}\rangle^2\right)\right\}=O(D^{-\tau}).$$
\end{definition}
Note that the "oracle" defined here is different from the traditional definition of "oracle", as we do not assume any "ground truth" model here. The developed method is data-driven,  and is not constrained with any model pre-specification. The $\tau$-oracle community structure may not be identifiable, however, this should not be viewed as a problem, because the main goal here is to find a reasonable community structure but not to find a pre-specified community structure.

Given a value of $\tau$ {specified by the oracle}, community structures are classified into three categories $\mathcal{H}_{\tau,-}$ (under-fitted), $\mathcal{H}_{\tau}$ and $\mathcal{H}_{\tau,+}$ (over-fitted), defined respectively as 
\begin{align*}
\mathcal{H}_{\tau,-}&=\left\{\bm{C}_\alpha\colon \sigma_{\alpha,D}=O(D^{-\beta_{1,\alpha}}),\beta_{1,\alpha}<\tau\right\},\\
\mathcal{H}_{\tau}&=\left\{\bm{C}_\alpha\colon \sigma_{\alpha,D}=O(D^{-\tau})\right\},\\
\mathcal{H}_{\tau,+}&=\left\{\bm{C}_\alpha\colon \sigma_{\alpha,D} =O(D^{-\beta_{2,\alpha}}),\beta_{2,\alpha}>\tau\right\},
\end{align*}

Define $\Delta\aleph^{\tau}_{\alpha,d}=\aleph \bm{C}_{\alpha,1:d}-\aleph\widetilde{\bm{C}}_{\tau,1:d}$. With Assumptions (1)--(6) (see the supplementary material), we now develop the theorem stated below, which illustrates the conditions under which a $\tau$-oracle community structure can be found with probability 1 asymptotically, and also essentially demonstrates how the selection of the tuning parameter $\lambda_N$ influences the efficiency of functional reconstruction of the resulting filt-fPCs. 
\begin{theorem}
\label{thm1}
Suppose that Assumptions (1)--(6) hold, and $\mathcal{H}_{\tau}\ne\emptyset$, if $\lambda_N$ satisfies the following conditions
\begin{align*}
&\max_{\alpha}\{M^{\beta_{1,\alpha}/\gamma}\Delta\aleph^\tau_{\alpha,D}\}\lambda_N\to0,	\qquad \bm{C}_\alpha\in\mathcal{H}_{\tau,-}\\
&M^{\tau/\gamma}\min_{\alpha}\{\Delta\aleph^\tau_{\alpha,D}\}\lambda_N\to\infty,\qquad  \bm{C}_\alpha\in\mathcal{H}_{\tau,+},
\end{align*}
the selected community structure $\widehat{\bm{C}}$ by minimizing (\ref{GIC}) is a $\tau$-oracle community structure asymptotically almost surely.
\end{theorem}


A drawback of the selection procedure based on the above GIC criterion is the high number of possible community structures if $G$ is large, making it very computationally costly to obtain the GIC values for all structures. To overcome this limitation, we propose another iterative procedure described below.

\subsubsection{Iterative selection of thresholds}
In the algorithm, the selection of thresholds determines the community structure.  We propose the following iterative GIC selection procedure. Denote the empirical filt-fPC score as
$Z_{vn,d}=\langle R^{(d-1)}_{vn},\hat{\phi}_d^{(c_{v,d})} \rangle,$ 
then the GIC value at layer $d$ is
$$\mbox{GIC}(\bm{C}_d)=-\sum_{v=1}^Gf_v\left\{N_v^{-1}\sum_{n=1}^{N_v}Z^2_{vn,d}\right\}+\kappa_N(d) \aleph \bm{C}_d,$$
where $\kappa_N(d)$ is a non-increasing function with respect to $d$, indicating more idiosyncratic features are to be obtained as $d$ increases. We propose to select the threshold $\tau_d$, such that the resulting $\widehat{\bm{C}}_d$ and $\{\hat{\phi}_d^{(c_{v,d})}(t)\colon v=1,\ldots,G\}$ minimize the above quantity. Suppose that the $(d-1)$-th threshold is $\tau_{d-1}$, then the $d$-th threshold is searched along the interval $[0,\tau_{d-1}]$ ({note that a common threshold can be employed for consecutive multiple, or even infinitely many, layers}). In order to minimize computational burden, the thresholds are selected from finite number of threshold candidates, where each candidate truncates at least one more edge than the larger candidates. 

\subsubsection{Selection of penalty and dimension}
{\color{black}It is noted that the number of tuning parameters $\{\kappa_N(d)\colon d=1,\ldots,D\}$ increases as $D$ diverges. To reduce the complexity of tuning parameter selection, our proposal is to employ some parametric form for $\kappa_N(d)$, e.g., $\kappa_N(d)=a d^{-b}$, $\kappa_N(d)=a b^{-d}$ or $\kappa_N(d)=a/(1+b^{d-u})$. A large-valued and slow-decaying sequence $\{\kappa_N(d)\colon d\ge1\}$ typically leads to a parsimonious but inefficient filt-fPC representation.} In principle, the dimension $D$ is selected so that the first $D$ filt-fPCs capture most variation (e.g., $90\%$) for each group. The values $a,b$ should be selected so that the resulting community structure is parsimonious (small cardinality) and meanwhile leads to efficient filt-fPCs in functional reconstruction. {\color{black}Since ordinary fPCs are optimal in functional reconstruction, we propose to compare the filt-fPCs and ordinary fPCs to check the efficiency. Specifically, we first specify some candidates of the tuning parameters, and then select $D$ so that the $D$-dimensional ordinary fPC representation approximate all groups of functions well, and select $a,b$ among the candidates so that the filt-fPCs obtained based on the $D$-layers community structure explains at least 90\% variance explained by the ordinary fPCs for each group with the minimal cardinality $\aleph \bm{C}_{1\colon D}$}. Clearly, the selection of $D$ and $\{\kappa_N(d)\colon d=1,\ldots,D\}$ is data driven.

Denote the community structure selected by the iterative GIC procedure as $\widehat{\bm{C}}_{{iter}}$. The following theorem demonstrates that, under some regularity conditions on $\{\kappa_N(d)\colon d\ge1\}$,  the iteratively selected community structure will not fall into the under-fitted class if the sample size is large enough, which theoretically guarantees the good performance of the proposed iterative procedure.
\begin{theorem}
\label{th2}
If Assumption (1)---(6) hold and $\lambda_N=\sum_{d\ge1}\kappa_{N}(d)$ satisfies the first condition in Theorem~\ref{thm1}, say, $\max_{\alpha}\{M^{\beta_{1,\alpha}/\gamma}\Delta\aleph^\tau_{\alpha,D}\}\lambda_N\to0,$ for $\bm{C}_\alpha\in\mathcal{H}_{\tau,-}.$ Additionally if $M^{-1/2}\sum_{d=1}^Dd^p\kappa_N^{-1}(d)\to0,$ then $\widehat{\bm{C}}_{iter}\notin \mathcal{H}_{\tau,-}$ asymptotically almost surely.
\end{theorem}
This theorem gives a uniform condition under which the iterative GIC selection procedure is guaranteed to produce decent community structure for efficient filt-fPC approximation when the sample size is large enough, regardless of the structure of $\widehat{\bm{C}}_{iter}$.

\section{Simulation Studies}
\label{s4} 
The goal here is to investigate the ability of the filt-fPCs to capture the underlying structure of multi-group functions. In the simulations, samples were generated from the following model $X_{vn}(t)=\sum_{d=1}^5\xi_{vn,d}\psi_{vd}(t),$ where $\psi_{vd}(t)$ are orthonormal basis across both $v$ and $d$. 500 functions were simulated for each of the 16 groups ($\Pi_1,\ldots,\Pi_{16}$). To obtain $\{\psi_{vd}(t)\colon v\ge1, d\ge1\}$, 22 functions were first randomly simulated with 23 Fourier basis functions $\{F_i(t)\colon i=1,\ldots,23\}$. Then the Gram-Schmidt process was applied to obtain 22 orthonormal basis functions $\{B_1(t),\ldots,B_{22}(t)\}$. The scores $\{\xi_{vn,d}\colon d=1,\ldots,5\}$ are independent and follow normal distribution $\mathcal{N}(0,1.2^{-d})$ for $v=1,\ldots,12$, and $\mathcal{N}(0,1.2^{d-6}),$ for group $v=13,\ldots,16$. The five basis functions employed to generate functions in each group are shown in Table \ref{t0}. The two halves are identical, but group 13--16 do not have the same covariance functions of group 5--8 as the scores follow different distributions. {\color{black} Typically, two groups $v,v'$ have similar covariance structure if 1) and basis functions used to generate the functions $\{\psi_{vd}\colon d=1,\ldots,5\}$ are similar and 2) the covariance structure of the scores $\{\xi_{vn,d}\colon d=1,\ldots,5\}$ are close to be proportional, e.g., $\mbox{cov}(\bm{\xi}_{nv})\approx \gamma\mbox{cov}(\bm{\xi}_{nv'})$, where $\gamma$ represents some positive constant.} 
\begin{table}[!h]	
\caption{Basis functions of different groups.}
	\begin{tabular}{|p{0.1in}<{\centering}|p{0.2in}<{\centering}p{0.2in}<{\centering}p{0.2in}<{\centering}p{0.2in}<{\centering}p{0.3in}<{\centering}|p{0.1in}<{\centering}|p{0.2in}<{\centering}p{0.2in}<{\centering}p{0.2in}<{\centering}p{0.2in}<{\centering}p{0.3in}<{\centering}|}
	\hline
	\hline
	\multicolumn{1}{|c|}{$v$} & \multicolumn{5}{c|}{\{$\phi_{vd}(t)\colon d=1,2,3,4,5$\}} &\multicolumn{1}{c|}{$v$} & \multicolumn{5}{c|}{\{$\phi_{vd}(t)\colon d=1,2,3,4,5$\}}\\
         \hline 
         1 & $B_1(t)$ &  $B_2(t)$ & $B_3(t)$ &  $B_4(t)$ &  $B_5(t)$ & $9$ &  $B_1(t)$ &  $B_2(t)$ & $B_3(t)$ &  $B_4(t)$ &  $B_5(t)$  \\
	 2 & $B_1(t)$ &  $B_2(t)$ & $B_3(t)$ &  $B_4(t)$ &  $B_6(t)$ & $10$ &  $B_1(t)$ &  $B_2(t)$ & $B_3(t)$ &  $B_4(t)$ &  $B_6(t)$ \\
	 3 & $B_1(t)$ &  $B_2(t)$ & $B_7(t)$ &  $B_8(t)$ &  $B_9(t)$ & $11$ &  $B_1(t)$ &  $B_2(t)$ & $B_7(t)$ &  $B_8(t)$ &  $B_9(t)$ \\
	 4 & $B_1(t)$ &  $B_2(t)$ & $B_7(t)$ &  $B_8(t)$ &  $B_{10}(t)$ & $12$ &  $B_1(t)$ &  $B_2(t)$ & $B_7(t)$ &  $B_8(t)$ &  $B_{10}(t)$ \\
	 5 & $B_1(t)$ &  $B_{11}(t)$ & $B_{12}(t)$ &  $B_{13}(t)$ &  $B_{14}(t)$ & $13$ &  $B_1(t)$ &  $B_{11}(t)$ & $B_{12}(t)$ &  $B_{13}(t)$ &  $B_{14}(t)$ \\
	 6 & $B_1(t)$ &  $B_{11}(t)$ & $B_{12}(t)$ &  $B_{15}(t)$ &  $B_{16}(t)$ & $14$ &  $B_1(t)$ &  $B_{11}(t)$ & $B_{12}(t)$ &  $B_{15}(t)$ &  $B_{16}(t)$ \\
	 7 & $B_1(t)$ &  $B_{11}(t)$ & $B_{17}(t)$ &  $B_{18}(t)$ &  $B_{19}(t)$ & $15$ &  $B_1(t)$ &  $B_{11}(t)$ & $B_{17}(t)$ &  $B_{18}(t)$ &  $B_{19}(t)$ \\
	 8 & $B_1(t)$ &  $B_{11}(t)$ & $B_{20}(t)$ &  $B_{21}(t)$ &  $B_{22}(t)$ & $16$ &  $B_1(t)$ &  $B_{11}(t)$ & $B_{20}(t)$ &  $B_{21}(t)$ &  $B_{22}(t)$ \\
	 \hline 
	 \hline 
	\end{tabular}
	
	\label{t0}
\end{table}

The iterative GIC criterion was employed to detect the community structure, and the penalty term follows the form $\kappa(d)=a d^{-b}$. The selected candidates for $a$ are 0.05, 0.1, 0.2, 0.3, 0.5, and for $b$ are 1, 1.1, 1.2, 1.3, 1.4. Here we used the ratio 
$$R=\sum_{v=1}^{16}\sum_{n=1}^{500}\|R^{(5)}_{vn}\|^2\bigg/\sum_{v=1}^{16}\sum_{n=1}^{500}\|X_{vn}\|^2$$
to evaluate the {reconstruction} performance of the estimated filt-fPCs. The corresponding ratio $R$ according to different pairs of $a,b$ are displayed in Table~\ref{rss}. {\color{black}Note that the reconstruction accuracy is improved as the penalty values decrease. This is because smaller values of penalty leads to more communities, which will increase the construction efficiency of filt-fPCs while scarificing the ability of explaining commonality.}  A reasonable selection of $a,b$ is $0.1,1.2$, {\color{black}since the reconstruction accuracy cannot be substantially improved with more communities. The total variation is sufficiently explained with a total of 5 layers since the functions are simulated with 5 orthonormal bases. The estimated filt-fPCs and the simulated basis functions, the average norm of reconstruction residuals $r^{(D)}_{vn}(t)=X_{vn}(t)-\sum_{d=1}^D\langle X_{nv},\psi_d^{(c_{v,d})}\rangle\psi_d^{(c_{v,d})}$, and the boxplots of the filt-fPC scores are given in the supplementary materials.} The selected community structure is displayed below:
\allowdisplaybreaks
\begin{align*}
\text{1st layer}:&\ (\Pi_1\text{--} \Pi_{16}).\\
\text{2nd layer}:&\ (\Pi_1\text{--} \Pi_4, \Pi_9\text{--} \Pi_{12});\ (\Pi_5\text{--} \Pi_{8}, \Pi_{13}\text{--}\Pi_{16});\\
\text{3rd layer}:&\ (\Pi_1,\Pi_2, \Pi_9,\Pi_{10});\ (\Pi_3,\Pi_4, \Pi_{11},\Pi_{12});\\
&\ (\Pi_{5},\Pi_{13});\ (\Pi_6,\Pi_{14});\ (\Pi_7,\Pi_{15});\ (\Pi_{8}, \Pi_{16}).\\
\text{4th layer}:&\ (\Pi_1,\Pi_2, \Pi_9,\Pi_{10});\ (\Pi_3,\Pi_4, \Pi_{11},\Pi_{12});\\
&\ (\Pi_{5},\Pi_{13});\ (\Pi_6,\Pi_{14});\ (\Pi_7,\Pi_{15});\ (\Pi_{8},\Pi_{16}).\\
\text{5th layer}:&\ (\Pi_1,\Pi_{9});\ (\Pi_2,\Pi_{10});\ (\Pi_3,\Pi_{11});\ (\Pi_{4},\Pi_{12});\\
&\ (\Pi_5);\ (\Pi_{6});\ (\Pi_{7});\ (\Pi_{8});\ (\Pi_{13});\ (\Pi_{14});\ (\Pi_{15});\ (\Pi_{16}).
\end{align*}

{\color{black}The first layer extracts the most common components driven by $B_1(t)$. Since $B_1(t)$ is shared by all groups, there is only one community in the first layer, and the second common components are drive by $B_2(t)$ and $B_{11}(t)$, and so on. Figure 1 in the supplementary materials shows that the estimated filt-fPCs are efficient in functional reconstruction. The obtained filt-fPCs are able to explain nearly 100\% variation for each group. In Figure 2 (supplementary materials), for some group (e.g., group 15, 16), the variance of filt-fPC scores are different from that of the simulated scores. This is not an issue, since the aim here is not to uncover the "ground truth" structure, but to find a structure which produces filt-fPCs efficient in functional reconstruction and explain commonality well. In Figure 3 (supplementary materials), it is clear that the estimated filt-fPCs are similar to the simulated basis functions. This also justifies the efficiency of filt-fPCs in functional reconstruction. The result is robust to the selection of the tuning parameters $a$, $b$.}

\begin{table}[ht]
\caption{$R$ values (\%) according to each pair of $a,b$, the value in the parentheses is the total number of distinct filt-fPCs.}
	\centering
	\begin{tabular}{|p{0.4in}<{\centering}|p{0.8in}<{\centering}|p{0.8in}<{\centering}|p{0.8in}<{\centering}|p{0.8in}<{\centering}|p{0.8in}<{\centering}|}
	\hline
	\hline
	\diagbox{$a$}{$b$} & 1& 1.1& 1.2& 1.3& 1.4\\
         \hline 
        0.05 & 0.041 (35) & 0.041 (35) & 0.040 (42) & 0.040 (42) & 0.040 (42) \\
	0.1 & 3.291 (21) & 0.037 (27)&0.037 (27)&0.041 (33)&0.041 (33)\\
	0.2 &3.291 (21)&3.291 (21)&3.291 (21)&3.291 (21)&0.037 (27)\\
	0.3 &5.089 (19)&3.291 (21)&3.291 (21)&3.291 (21)&3.291 (21)\\
	0.5 &18.27 (13)&9.470 (17)&9.470 (17)&5.089 (19)&5.089 (19)\\
	 \hline 
	 \hline 
	\end{tabular}
	\label{rss}
\end{table}


{\color{black} For comparison, we also implemented the partial common functional principal component analysis (PCfPCA), and estimate the common fPCs with the semi-parametric method proposed by Wang et al.~(2019) \cite{wang2019semiparametric}. The number of common fPCs (denoted by $\#CPC$) takes value in $1, 2, \ldots, 5$ (The PCfPC model degenerate to a common functional principal component model when there are 5 common fPCs). Specifically, in PCfPCA, we assume that
\begin{equation*}
X_{vn}(t)=\left\{
\begin{array}{ccl}
\sum\limits_{d=1}^{\#CPC}\xi_{vn,d}\phi_d+\sum\limits_{d=\#CPC+1}^{5}\xi_{vn,d}\phi_{vd},& & if\ \#CPC<5,\\
\sum\limits_{d=1}^{5}\xi_{vn,d}\phi_d,& & if\ \#CPC=5.
\end{array} \right. 
\end{equation*}
where $\langle\phi_d,\phi_{d'}\rangle=0$ for $d\ne d'$, $\langle\phi_d,\phi_{vd'}\rangle=0$ for $d'>d$, and $\phi_d,\phi_{d'}$ are normalized functions.

The corresponding $R$ values are shown in Table \ref{rss_PCPCA}. Note that, only when there is one common fPC, the reconstruction accuracy is decent. This is because there is only one common fPC across all groups, and other common fPCs are only present among partial groups. Clearly, PCfPCA is not sufficient to explain such complex commonality structure. The performance of filt-fPCA when PCfPC model is correct is still decent, and the results under this setting are in the supplementary materials.}

\begin{table}[ht]
\caption{$R$ values (\%) of PCfPCA with different number of common fPCs, and the value in the parentheses are the number of distinct fPCs.}
	\centering
	\begin{tabular}{|p{0.5in}<{\centering}|p{0.8in}<{\centering}|p{0.8in}<{\centering}|p{0.8in}<{\centering}|p{0.8in}<{\centering}|p{0.8in}<{\centering}|}
	\hline
	\hline
	\#CPC & 1& 2& 3& 4& 5\\
         \hline 
        $R$ & 0.006 (65) & 7.417 (50) &  14.62 (35) & 28.94 (20) & 45.01 (5) \\
	 \hline 
	 \hline 
	\end{tabular}
	
	\label{rss_PCPCA}
\end{table}

{\color{black}In CfPCA/PCfPCA, the type of commonality structure is fixed (all groups share the same set of common fPCs). However, such simple model is not sufficient to explain complex commonality structure. Comparatively, in filt-fPCA we aim to find a reasonable commonality structure, which solves the limitation of CfPCA/PCfPCA.}

\section{The Analysis of Rat Brain Local Field Potentials}
\label{s5}
Synchrony widely exists in brain signals, and is an important measure of coordination of brain. A suddenly increased scale of synchrony can indicate a rapidly emerging response from an extreme shock such as stroke. Here, we applied the proposed filt-fPCA to analyze the changes in the synchrony structure of LFPs collected from a rat brain across 32 recorded regions.
\subsection{Data processing and visualization}
\label{s5.1}
The LFPs were bandpass filtered at (0, {\color{black}50}] Hertz and segmented into 1-second epochs. The same procedure can also be employed for other frequency bands but we did not pursue it here. In the situation where structural breaks in the covariance structure are also of major concern, the first step is to detect these breakpoints (e.g., Jiao, Frostig and Ombao (2020) \cite{jiao2022break}), and then apply the proposed filt-fPCA method to each local quasi-stationary sub-sequence segmented by the detected break points. Here we considered the overall difference of variation pattern and hidden community structure between the pre-occlusion and post-occlusion epochs, so we conducted a global analysis for both phases.

Here we consider a maximal of 25 layers. Visualization reveals the occasional occurrence of irregular extreme values. Therefore, to stabilize the variance, we applied the square root transformation to the values of trajectories. In addition, outlier epochs were removed from each tetrode under both phases, where outlier epochs in each group are defined as those of which the $l^2$-norm is beyond the interval $[Q_1 - 1.5\times \mbox{IQR},Q_3 + 1.5\times \mbox{IQR}]$. Here $\mbox{IQR}=Q_3-Q_1$ and $Q_1, Q_3$ are the first and third quantile of the $l_2$-norm of the epoch trajectories. 
 
Two networks $(N,E,\bm{\omega}_1)$, \ $(N,E,\bm{\omega}_2)$ were constructed for the pre-occlusion and post-occlusion epochs separately. The node set $N$ has 32 nodes representing the 32 tetrodes, and the edge set $E$ is complete initially. Figure~\ref{f9} displays the edge weights $\omega^{(k)}_{ij}=\|\mathcal{C}^{(k)}_{i}-\mathcal{C}^{(k)}_{j}\|_\mathcal{S},\ k=1,2,\ i,j=1,\ldots,32$. Here, $k=1$ indicates the pre-occlusion phase and $k=2$ refers to the post-occlusion phase. Figure \ref{sw} shows the average weights of edges adjacent to each node, defined as $\sum_{j=1}^{32}\omega^{(k)}_{ij}/32$, $i=1,2,\ldots,32$. 
\begin{figure}[ht]
\center
\includegraphics[scale=0.13]{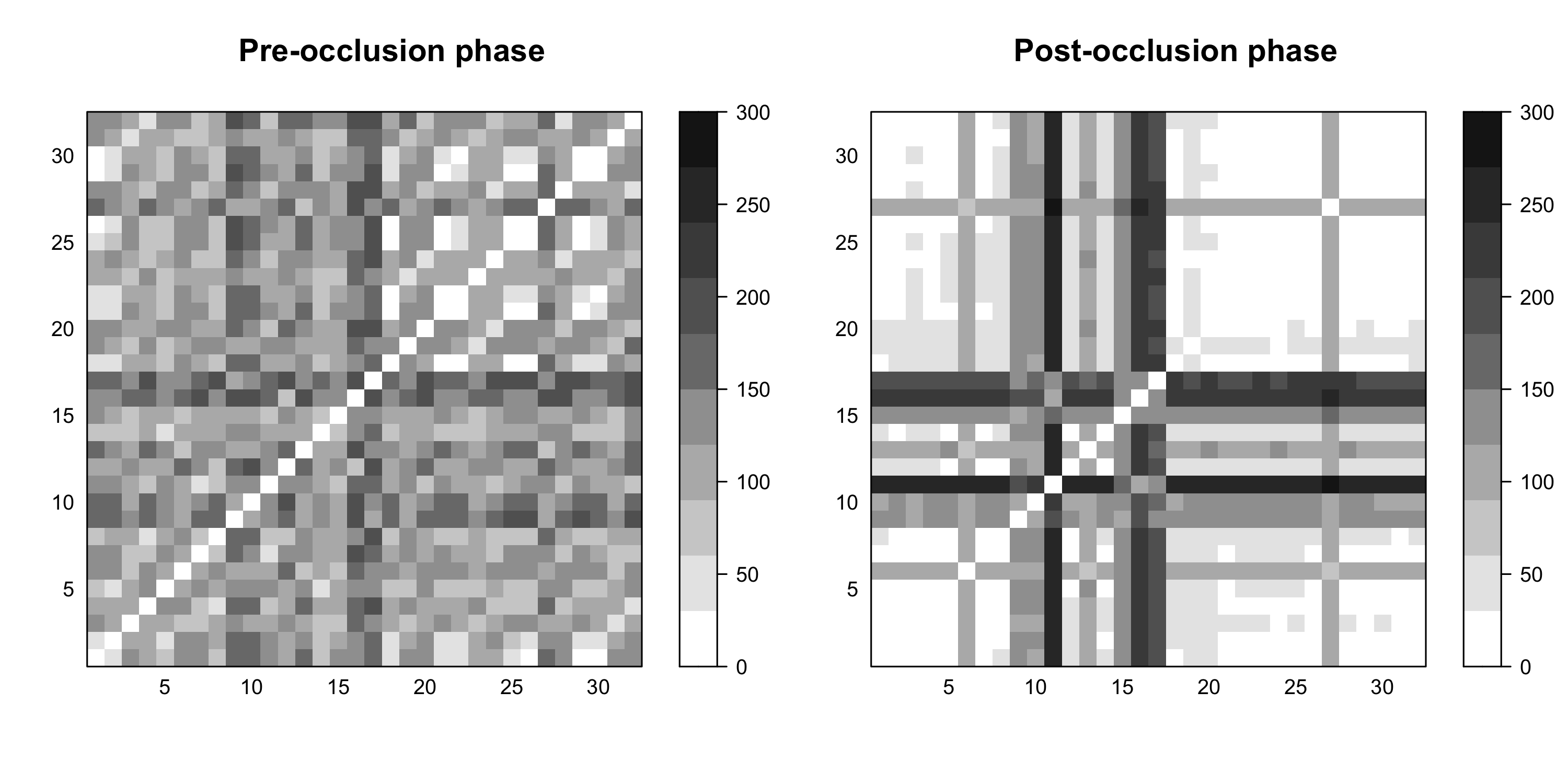}
\caption{$\|\mathcal{C}^{(k)}_{i}-\mathcal{C}^{(k)}_{j}\|_\mathcal{S}$, $i,j=1,\ldots,32$, $k=1,2$.}
\label{f9}
\end{figure}
\begin{figure}[ht]
\center
\includegraphics[scale=0.13]{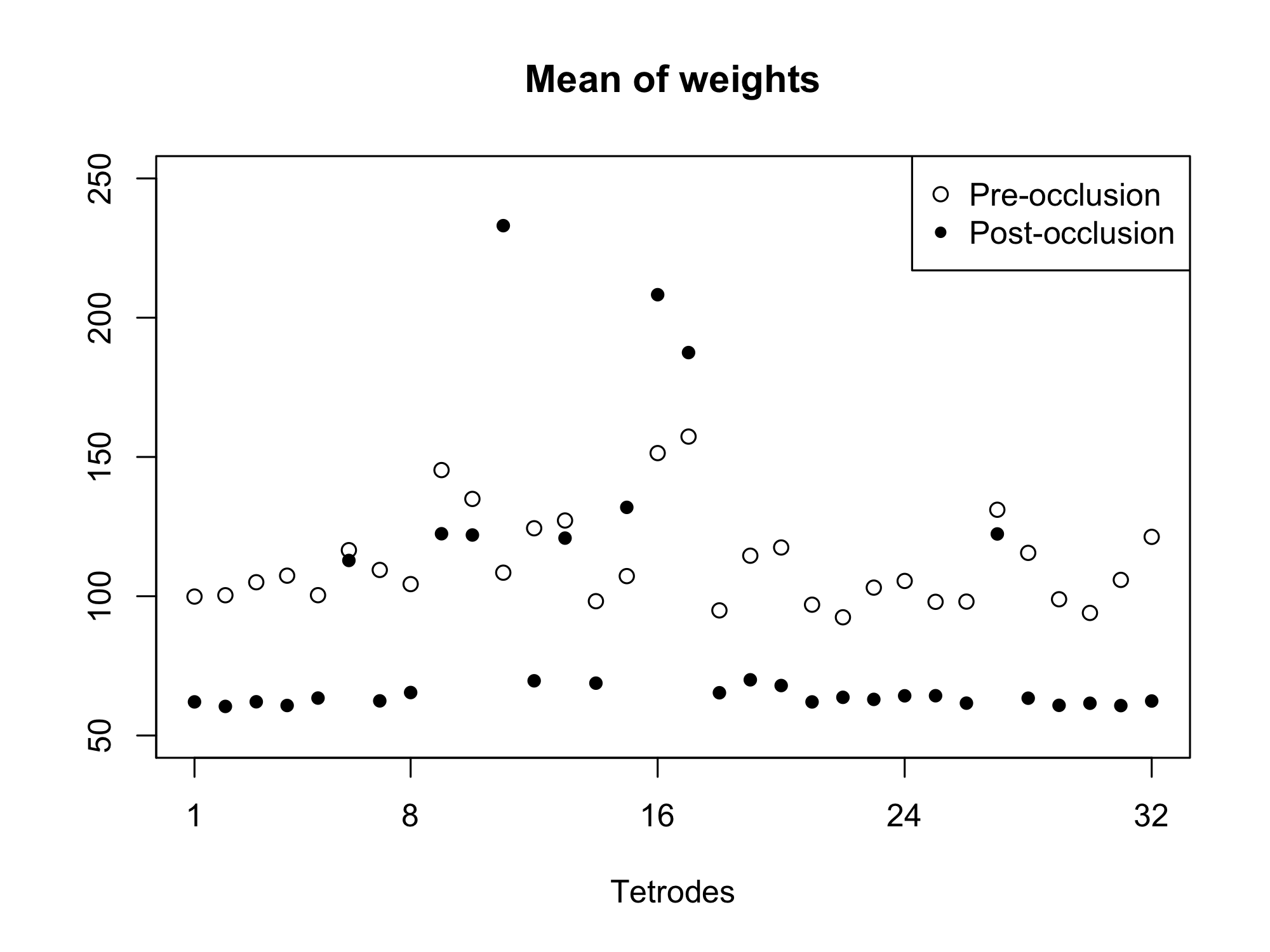}
\caption{Average weights of edges adjacent to each node} 
\label{sw}
\end{figure}
In Figure \ref{f9} and \ref{sw}, after the occlusion onset, the epoch trajectories from most of the 32 tetrodes display more similar variation patterns due to the increased  larger-scale signal-synchrony. Therefore, a more parsimonious filt-fPC representation is expected for the post-occlusion phase. The weight matrix changes substantially after the occlusion onset, making it necessary to implement filt-fPCA to the two phases separately. We also justified the necessity of comparison through a permutation test of weight matrix, and the details can be found in the supplementary materials.

\subsection{Community detection and filt-fPCs}
\label{s5.3}
The iterative GIC selection procedure was employed to select the community structure. To conduct a fair comparison of community structures, we set $\kappa(d)=0.007\times d^{-1.4}$ for both phases. {\color{black}We chose this value because the resulting first 25 layers filt-fPCs explain at least 95\% variation that 25 group-wise ordinary fPCs explain for each tetrode with the minimal cardinality.}  

Figure~\ref{precom} and \ref{postcom} show the communities (first 8 layers) of the two phases, where the points (representing tetrodes, displayed in the same order as in the experiment) with the same color and shape are in the same community. Clearly, after the occlusion onset, most tetrodes were clustered in the same community, leading to a much more parsimonious filt-fPC representation. The total number of distinct communities drops from 331 to 202 after the artery occlusion (25 layers). One interpretation here is that an extreme event such as the sudden lack of oxygen delivered caused the neurons to respond in a similar manner. It is interesting that this phenomenon is also observed in financial data, i.e., a severe drop in the market elicits similar and synchronized behavior in stocks. The first five layers' filt-fPCs of pre-occlusion and post-occlusion trajectories are presented in Figure \ref{filt-fpc}. {\color{black}It is noted that the first few filt-fPCs explain low-frequency oscillations, and this coincides with the pre-knowledge that the synchrony is mainly driven by low frequencies (see Wann (2017) \cite{wann2017large}).}
\begin{figure}[H]
\center
\includegraphics[scale=0.14]{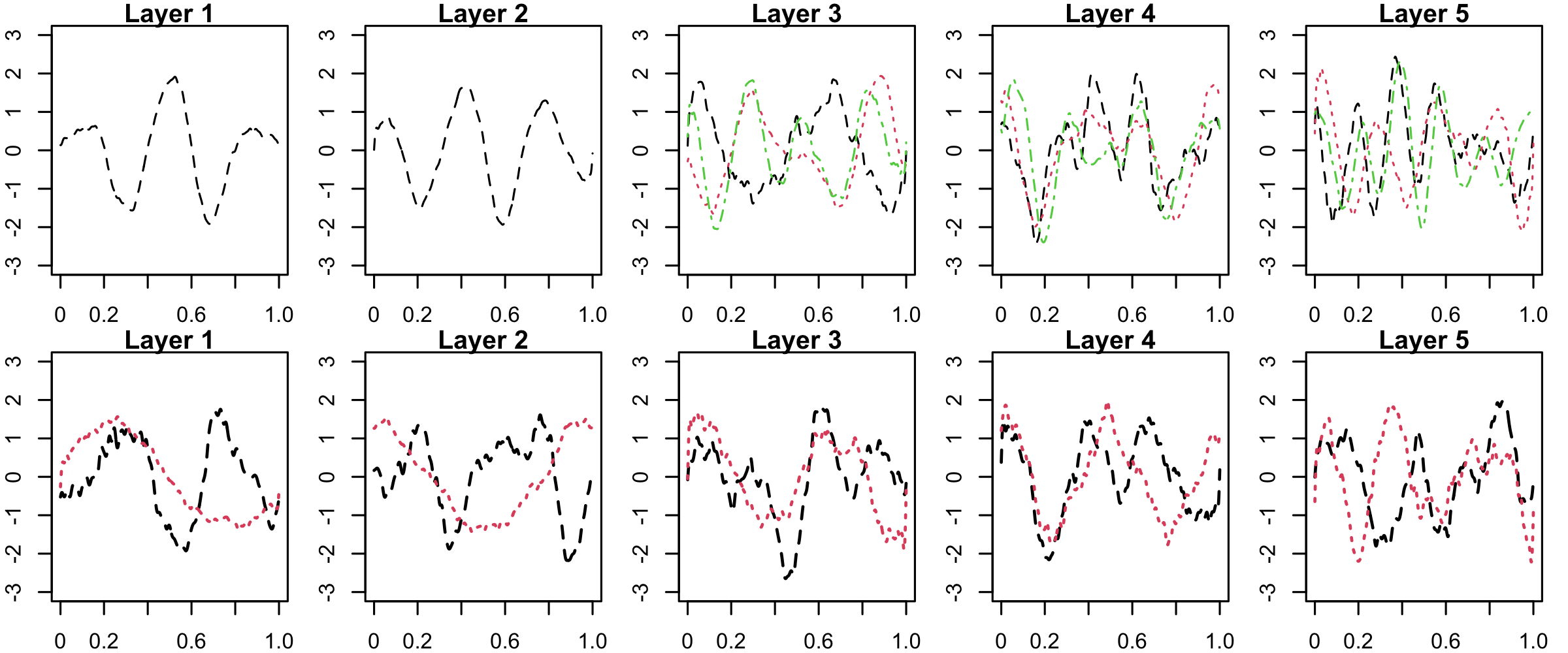}
\caption{Estimated filt-fPCs of the first four layers. The upper 5 figures pertain to the pre-occlusion phase, and the lower 5 figures pertain to the post-occlusion phase.}
\label{filt-fpc}
\end{figure}
\begin{figure}[ht]
\center
\includegraphics[scale=0.15]{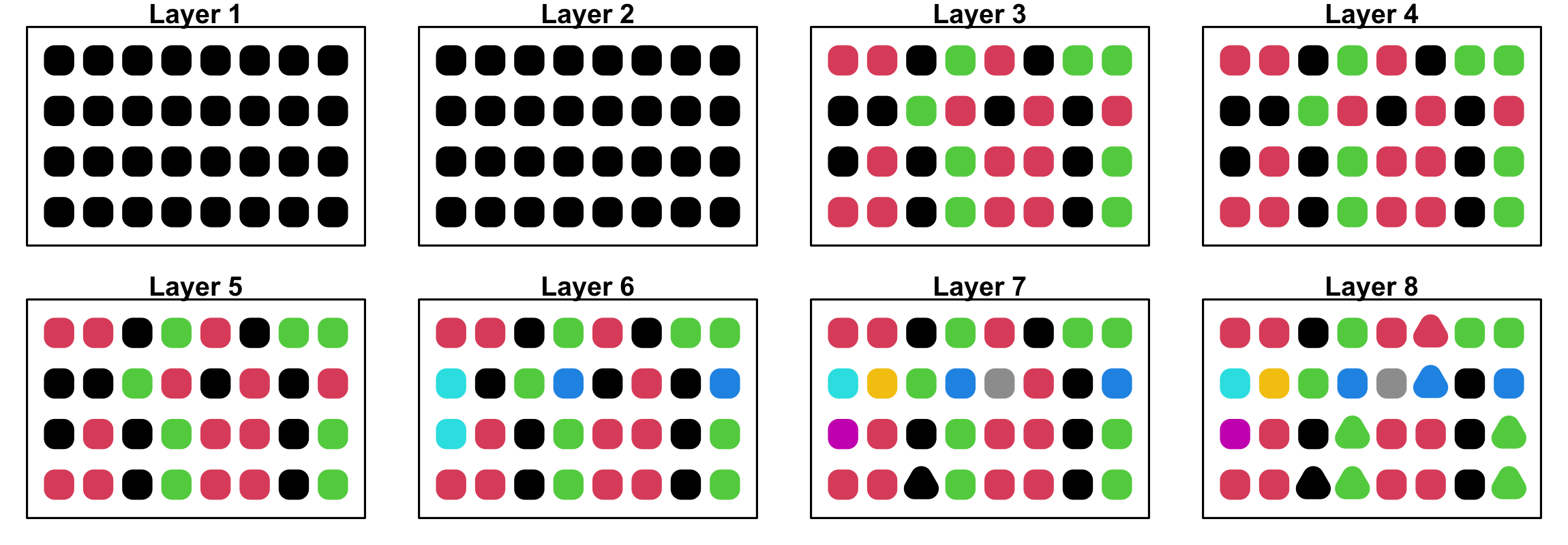}
\caption{Community structures of pre-occlusion phase}
\label{precom}
\end{figure}
\begin{figure}[ht]
\includegraphics[scale=0.15]{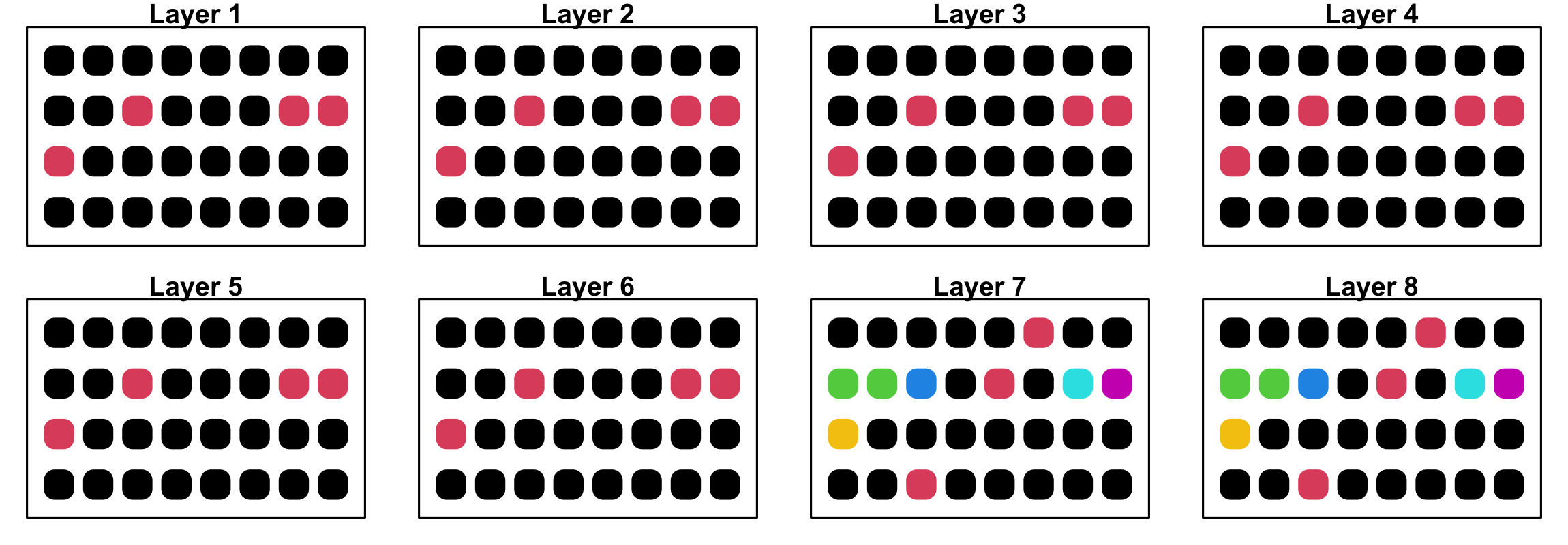}
\caption{Community structures of post-occlusion phase}
\label{postcom}
\end{figure}
{\color{black}We found that, before the occlusion, the tetrode (1, 2, 18, 21, 22, 25, 26, 29, 30), (24, 28, 32), (3, 31), (4,8), (7,11), and (19,23) are consistently assigned in the same community respectively across the 25 layers, and after the occlusion, the tetrode (1-- 5, 7, 8, 12, 14, 29), (18 -- 26, 28, 30--32) are consistently assigned in the same community respectively across the 25 layers, the brain locations over which tetrodes are assigned in the same community are potentially strongly inter-connected.}

\subsection{Reconstruction efficiency of filt-fPCs}
 {\color{black}To show the efficiency of the obtained filt-fPCs in functional reconstruction, here we check the difference between the reconstruction of the filt-fPCs and that of the partial common fPCs. Specifically, we compute the following value to evaluate the reconstruction efficiency.
$$e^{(k)}_{v,D}=\frac{\left\|\frac{1}{N_v}\sum_{n=1}^{N_v}\sum_{d=1}^{D}\langle X^{(k)}_{vn},\hat{\phi}_{vd}\rangle\right\|^2-\left\|\frac{1}{N_v}\sum_{n=1}^{N_v}\sum_{d=1}^{D}\langle X^{(k)}_{vn},\hat{\phi}_{d}^{(c_{v,d})}\rangle\right\|^2}{\left\|\frac{1}{N_v}\sum_{n=1}^{N_v}\sum_{d=1}^{25}\langle X^{(k)}_{vn},\hat{\phi}_{vd}\rangle\right\|^2},$$
for $D=1,\ldots,25$, where $\hat{\phi}_{d}^{(c_{v,d})}$ denotes filt-fPC or fPC in a PCfPC model (obtained by the approach \cite{wang2019semiparametric}). By the method \cite{wang2019semiparametric}, there are 12 and 22 common fPCs detected for the pre-occlusion and post-occlusion phase respectively. Since group-wise ordinary fPCs are optimal in functional reconstruction, they serve as the baseline of comparison, and a small value of $e^{(k)}_{v,D}$ indicates that the $D$-dimensional filt-fPCs (or PCfPCs) representation is close to the $D$-dimensional ordinary fPC representation.
The average values of $e^{(k)}_{v,D}$ across all tetrodes are displayed in Figure \ref{average-norm}, which shows the overall better reconstruction performance of filt-fPCs.
\begin{figure}[ht]
\centering
\includegraphics[scale=0.14]{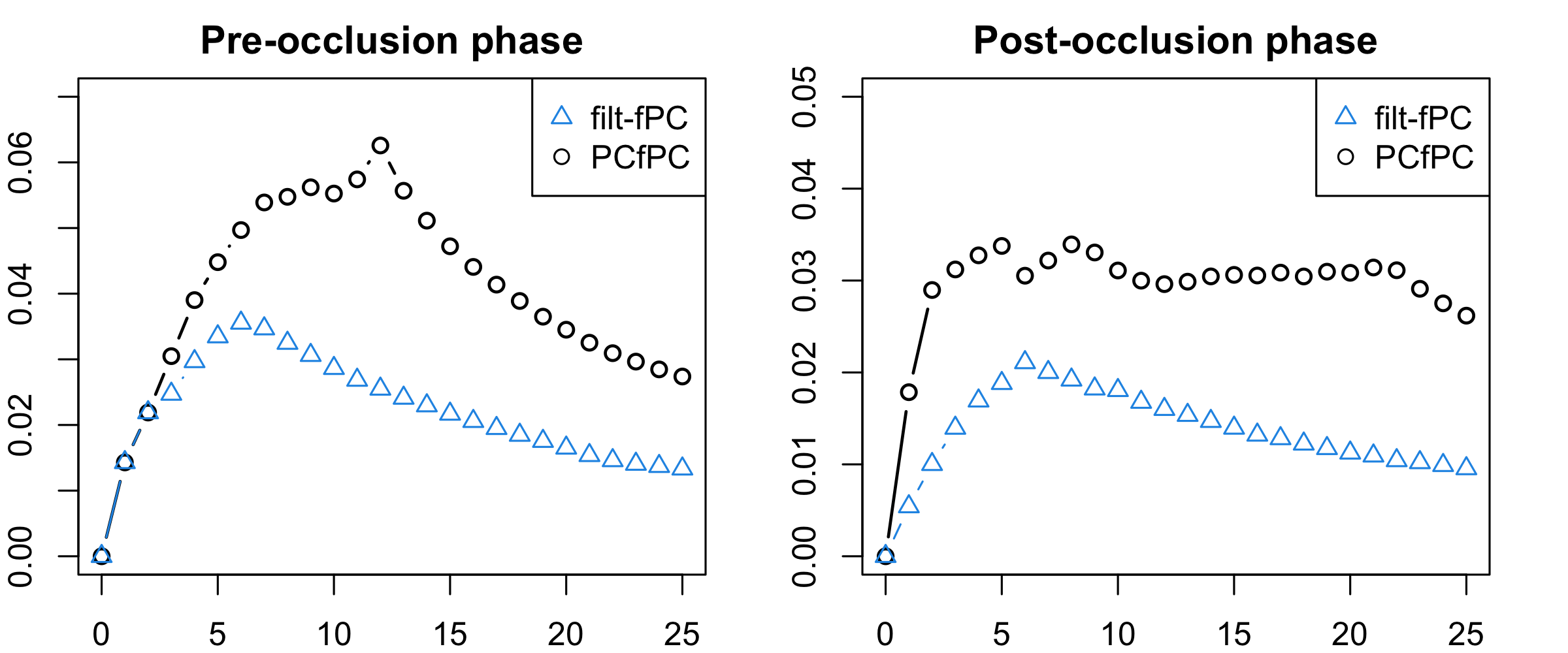}
\caption{Average $\{e^{(k)}_{v,D}\colon D=1,\ldots,25\}$ across all tetrodes.}
\label{average-norm}
\end{figure}

The variance of the filt-fPC scores are displayed in Figure 4 and 5 in the supplementary materials. It is noted that the spectral distribution of tetrode 6, 9, 10, 13, 17, 27 is flat, while variation of epochs collected from other tetrodes are mainly explained by the first few filt-fPCs. The tetrode-wise values of $e^{(k)}_{v,D}$ are displayed in Figure 6 and 7 in the supplementary materials. Clearly, for all tetrodes, $e^{(k)}_{v,25}<0.05$, and this means that the first 25 filt-fPCs can explain at least 95\% variation explained by the group-wise ordinary fPCs. This justifies the efficiency of the selected filt-fPCs in terms of functional reconstruction. For tetrode 6, 9, 10, 13, 17, 27, the PCfPC model leads to more than 10\% loss of variation compared with group-wise ordinary fPCs. In addition, for most $D=1,\ldots, 25$, the $e^{(k)}_{v,D}$ value of PCfPC model is higher than that of filt-fPC model. All of these findings justifies the superiority of functional reconstruction of filt-fPCs. That is because the variation pattern of epochs collected from these tetrodes are substantially different from other tetrodes and it is not advantageous to extract the common principal components for all the tetrodes jointly. In filt-fPCA, these tetrodes are separated from the others at the first few layers, and thus the obtained filt-fPCs are more efficient. In addition, the restriction of PCfPC model fails to reveal the commonality structure across tetrodes in a sophisticated manner, in other words, PCfPCA cannot reveal different levels of synchrony across groups. }


\section{Conclusions}
\label{s6}
Local field potentials provide information about brain function. They are collected from multiple tetrodes inserted on a pre-arranged patch on the cortex. The trajectories collected from different tetrodes simultaneously can be considered as multi-group functional data. Synchrony of different tetrodes potentially indicates functional connectivity of different regions of brain and a suddenly increased scale  of spontaneous neuronal synchrony may antecede neuronal activity impairments in ischemic studies. The filt-fPC analysis provides a novel and efficient way to extract and quantify the multi-layer synchrony structure of multi-tetrode LFP recordings by employing filtrated common functional principal components. We developed a data-driven algorithm to find the filtrated common functional principal components. Specifically, we first specify a tree-structured community structure for the weighted network, established from data to measure the similarity of covariance structures of different tetrodes of epochs, and then find the common filt-fPCs for every community. The application of filt-fPCA to the local field potentials not only reveal the large-scale synchrony phenomenon after the occlusion onset, but also quantify the changes of synchrony. The method is developed {\it not} only for the LFP data described in this paper, but for all kinds of multi-group functional data, such as longitudinal functional data, multivariate functional data and spatial-temporal data. 

The filt-fPCA have several advantages: (1.) The method is data driven, can be implemented without prior-knowledge, and is not constrained by model pre-specification, making it suitable for any complex case. (2.) The method is able to extract the common variation patterns of different groups of functions in a novel "multi-resolution" manner, and the obtained filt-fPCs are efficient in functional reconstruction. (3.) {The method can be applied to both balanced and unbalanced design}. (4.) The filtrated functional principal components are orthogonal to each other for each group, leading to a concise basis representation. Extending filtration techniques to functional linear models will be pursued as future work.

\bibliography{filt-fPC-arkiv-version}

@article{bali2011robust,
  title={Robust functional principal components: A projection-pursuit approach},
  author={Bali, Juan Lucas and Boente, Graciela and Tyler, David E and Wang, Jane-Ling},
  journal={The Annals of Statistics},
  volume={39},
  number={6},
  pages={2852--2882},
  year={2011},
  publisher={Institute of Mathematical Statistics}
}

@article{hall2006properties1,
  title={On properties of functional principal components analysis},
  author={Hall, Peter and Hosseini-Nasab, Mohammad},
  journal={Journal of the Royal Statistical Society: Series B (Statistical Methodology)},
  volume={68},
  number={1},
  pages={109--126},
  year={2006},
  publisher={Wiley Online Library}
}

@article{hall2006properties2,
  title={Properties of principal component methods for functional and longitudinal data analysis},
  author={Hall, Peter and M{\"u}ller, Hans-Georg and Wang, Jane-Ling},
  journal={The annals of statistics},
  pages={1493--1517},
  year={2006},
  publisher={JSTOR}
}

@article{yao2006penalized,
  title={Penalized spline models for functional principal component analysis},
  author={Yao, Fang and Lee, Thomas CM},
  journal={Journal of the Royal Statistical Society: Series B (Statistical Methodology)},
  volume={68},
  number={1},
  pages={3--25},
  year={2006},
  publisher={Wiley Online Library}
}

@article{jiang2010covariate,
author = {Ci-Ren Jiang and Jane-Ling Wang},
title = {{Covariate adjusted functional principal components analysis for longitudinal data}},
volume = {38},
journal = {The Annals of Statistics},
number = {2},
publisher = {Institute of Mathematical Statistics},
pages = {1194 -- 1226},
year = {2010},
doi = {10.1214/09-AOS742},
}

@article{yao2007functional,
  title={Functional principal component analysis for longitudinal and survival data},
  author={Yao, Fang},
  journal={Statistica Sinica},
  pages={965--983},
  year={2007},
  publisher={JSTOR}
}

@article{chen2017modelling,
  title={Modelling function-valued stochastic processes, with applications to fertility dynamics},
  author={Chen, Kehui and Delicado Useros, Pedro Francisco and M{\"u}ller, Hans-Georg},
  journal={Journal of the Royal Statistical Society. Series B, Statistical Methodology},
  volume={79},
  number={1},
  pages={177--196},
  year={2017}
}

@article{zhang2010regularization,
  title={Regularization parameter selections via generalized information criterion},
  author={Zhang, Yiyun and Li, Runze and Tsai, Chih-Ling},
  journal={Journal of the American Statistical Association},
  volume={105},
  number={489},
  pages={312--323},
  year={2010},
  publisher={Taylor \& Francis}
}

@article{chiou2014multivariate,
  title={Multivariate functional principal component analysis: A normalization approach},
  author={Chiou, Jeng-Min and Chen, Yu-Ting and Yang, Ya-Fang},
  journal={Statistica Sinica},
  pages={1571--1596},
  year={2014},
  publisher={JSTOR}
}

@article{jiao2022break,
  title={Break Point Detection for Functional Covariance},
  author={Jiao, Shuhao and Frostig, RD and Ombao, Hernando},
  journal={arXiv:2006.13887},
  year={2022}
}

@article{chen2012modeling,
  title={Modeling repeated functional observations},
  author={Chen, Kehui and M{\"u}ller, Hans-Georg},
  journal={Journal of the American Statistical Association},
  volume={107},
  number={500},
  pages={1599--1609},
  year={2012},
  publisher={Taylor \& Francis}
}

@article{nishii1984asymptotic,
  title={Asymptotic properties of criteria for selection of variables in multiple regression},
  author={Nishii, Ryuei},
  journal={The Annals of Statistics},
  pages={758--765},
  year={1984},
  publisher={JSTOR}
}

@article{flury1987two,
  title={Two generalizations of the common principal component model},
  author={Flury, Bernhard K},
  journal={Biometrika},
  volume={74},
  number={1},
  pages={59--69},
  year={1987},
  publisher={Oxford University Press}
}

@article{wang2019semiparametric,
  title={Semiparametric Partial Common Principal Component Analysis for Covariance Matrices},
  author={Wang, Bingkai and Luo, Xi and Zhao, Yi and Caffo, Brain},
  journal={bioRxiv},
  pages={808527},
  year={2019},
  publisher={Cold Spring Harbor Laboratory}
}

@article{benko2009common,
  title={Common functional principal components},
  author={Benko, Michal and H{\"a}rdle, Wolfgang and Kneip, Alois },
  journal={The Annals of Statistics},
  volume={37},
  number={1},
  pages={1--34},
  year={2009},
  publisher={Institute of Mathematical Statistics}
}

@article{schott1999partial,
  title={Partial common principal component subspaces},
  author={Schott, James R},
  journal={Biometrika},
  volume={86},
  number={4},
  pages={899--908},
  year={1999},
  publisher={Oxford University Press}
}

@article{crainiceanu2011population,
  title={Population value decomposition, a framework for the analysis of image populations},
  author={Crainiceanu, Ciprian M and Caffo, Brian S and Luo, Sheng and Zipunnikov, Vadim M and Punjabi, Naresh M},
  journal={Journal of the American Statistical Association},
  volume={106},
  number={495},
  pages={775--790},
  year={2011},
  publisher={Taylor \& Francis}
}

@article{flury1984common,
  title={Common principal components in k groups},
  author={Flury, Bernhard N},
  journal={Journal of the American Statistical Association},
  volume={79},
  number={388},
  pages={892--898},
  year={1984},
  publisher={Taylor \& Francis}
}

@article{happ2018multivariate,
  title={Multivariate functional principal component analysis for data observed on different (dimensional) domains},
  author={Happ, Clara and Greven, Sonja},
  journal={Journal of the American Statistical Association},
  volume={113},
  number={522},
  pages={649--659},
  year={2018},
  publisher={Taylor \& Francis}
}

@article{berrendero2011principal,
  title={Principal components for multivariate functional data},
  author={Berrendero, Jos{\'e} R and Justel, Ana and Svarc, Marcela},
  journal={Computational Statistics \& Data Analysis},
  volume={55},
  number={9},
  pages={2619--2634},
  year={2011},
  publisher={Elsevier}
}

@incollection{greven2011longitudinal,
  title={Longitudinal functional principal component analysis},
  author={Greven, Sonja and Crainiceanu, Ciprian and Caffo, Brian and Reich, Daniel},
  booktitle={Recent Advances in Functional Data Analysis and Related Topics},
  pages={149--154},
  year={2011},
  publisher={Springer}
}

@article{ramsay2004functional,
  title={Functional data analysis},
  author={Ramsay, James O and Silverman, Bernard W},
  journal={Encyclopedia of Statistical Sciences},
  volume={4},
  year={2004},
  publisher={Wiley Online Library}
}

@article{jacques2014model,
  title={Model-based clustering for multivariate functional data},
  author={Jacques, Julien and Preda, Cristian},
  journal={Computational Statistics \& Data Analysis},
  volume={71},
  pages={92--106},
  year={2014},
  publisher={Elsevier}
}

@article{kayano2009functional,
  title={Functional principal component analysis via regularized Gaussian basis expansions and its application to unbalanced data},
  author={Kayano, Mitsunori and Konishi, Sadanori},
  journal={Journal of Statistical Planning and Inference},
  volume={139},
  number={7},
  pages={2388--2398},
  year={2009},
  publisher={Elsevier}
}

@article{di2009multilevel,
  title={Multilevel functional principal component analysis},
  author={Di, Chong-Zhi and Crainiceanu, Ciprian M and Caffo, Brian S and Punjabi, Naresh M},
  journal={The annals of applied statistics},
  volume={3},
  number={1},
  pages={458},
  year={2009},
  publisher={NIH Public Access}
}

@article{di2014multilevel,
  title={Multilevel sparse functional principal component analysis},
  author={Di, Chongzhi and Crainiceanu, Ciprian M and Jank, Wolfgang S},
  journal={Stat},
  volume={3},
  number={1},
  pages={126--143},
  year={2014},
  publisher={Wiley Online Library}
}

@phdthesis{wann2017large,
  title={Large-scale spatiotemporal neuronal activity dynamics predict cortical viability in a rodent model of ischemic stroke},
  author={Wann, Ellen Genevieve},
  year={2017},
  school={UC Irvine}
}

@article{lock2013joint,
  title={Joint and individual variation explained (JIVE) for integrated analysis of multiple data types},
  author={Lock, Eric F and Hoadley, Katherine A and Marron, James Stephen and Nobel, Andrew B},
  journal={The annals of applied statistics},
  volume={7},
  number={1},
  pages={523},
  year={2013},
  publisher={NIH Public Access}
}

@article{feng2018angle,
  title={Angle-based joint and individual variation explained},
  author={Feng, Qing and Jiang, Meilei and Hannig, Jan and Marron, JS},
  journal={Journal of multivariate analysis},
  volume={166},
  pages={241--265},
  year={2018},
  publisher={Elsevier}
}
\bibliographystyle{agsm}

\end{document}